\documentclass[preprint,aps]{revtex4}
\usepackage{amsmath}
\usepackage{epsfig}
\usepackage{graphicx}
\usepackage{hyperref}
\usepackage{url}

\newcommand{\beq}{\begin{equation}}
\newcommand{\eeq}{\end{equation}}
\newcommand{\bea}{\begin{eqnarray}}
\newcommand{\eea}{\end{eqnarray}}
\newcommand{\ba}{\begin{array}}
\newcommand{\ea}{\end{array}}
\newcommand{\Msun}{{\rm M}_{\odot}}
\newcommand{\ie}{i.~e.~}

\def\leq{\,\raise 0.4ex\hbox{$<$}\kern -0.8em\lower 0.62ex\hbox{$-$}\,}
\def\geq{\,\raise 0.4ex\hbox{$>$}\kern -0.8em\lower 0.62ex\hbox{$-$}\,}
\def\pm{\,\raise 0.4ex\hbox{$+$}\kern -0.8em\lower 0.62ex\hbox{$-$}\,}
\def\lsim{\,\raise 0.4ex\hbox{$<$}\kern -0.8em\lower 0.62ex\hbox{$\sim$}\,}
\def\gsim{\,\raise 0.4ex\hbox{$>$}\kern -0.8em\lower 0.62ex\hbox{$\sim$}\,}
\def\propsim{\,\raise 0.4ex\hbox{$\propto$}\kern -0.8em\lower 0.62ex\hbox{$\sim$}\,}

\begin{document}

\title{Electromagnetic extraction of energy from black hole-neutron star binaries}

\date{\today}

\author{Sean T. McWilliams${}^{1,2,3}$ and Janna Levin${}^{1,3}$}
\affiliation{${}^{1}$Institute for Strings, Cosmology and Astroparticle Physics (ISCAP), Columbia University, New York, NY 10027}
\affiliation{${}^{2}$Physics Department, Princeton University, Princeton, NJ 08544}
\affiliation{${}^{3}$Department of Physics and Astronomy, Barnard College, New York, NY 10027}
\email{sean@astro.columbia.edu}

\begin{abstract}
{\bf
The coalescence of black hole-neutron star binaries is expected to be a principal source of gravitational waves for the next
generation of detectors, Advanced LIGO and Advanced Virgo.  Ideally, these and other gravitational wave sources would have a distinct
electromagnetic counterpart, as significantly more information could be gained through two separate channels.
In addition, since these detectors will probe distances with non-negligible redshift, a coincident observation
of an electromagnetic counterpart to a gravitational wave signal would facilitate a novel measurement of dark energy \cite{Holz:2005df}.  
For black hole masses not much larger than the
neutron star mass, the tidal disruption and subsequent accretion of the neutron star by the black hole provides one
avenue for generating an electromagnetic counterpart \cite{Lattimer}.  However, in this work, we demonstrate that, 
for all black hole-neutron star
binaries observable by Advanced LIGO/Virgo,
the interaction of the black hole with the magnetic field of the neutron star will drive a Poynting flux.  This Poynting
flux generates synchrotron/curvature radiation as the electron-positron plasma in the neutron star magnetosphere is accelerated, and 
thermal radiation
as the plasma is focused onto the neutron star magnetic poles, creating a ``hot spot'' on the neutron
star surface.  This novel effect will generate copious luminosity, comparable to supernovae and active galactic nuclei, so that 
black hole-neutron star coalescences detectable with gravitational waves by Advanced LIGO/Virgo
could also potentially be detectable electromagnetically.
}
\end{abstract}

\maketitle

When a neutron star is tidally disrupted by a stellar-mass black hole, the system will
release substantial amounts of energy as the remnant material is accreted.  Along with neutron star-neutron star binary mergers,
these events are a leading candidate for driving short $\gamma$-ray 
bursts (GRBs), which persist for less than $\sim 2$ seconds, yet can release more energy
in that time than most galaxies will emit in an entire year.
Apart from being the most luminous occurence
in the Universe, these short GRBs are potentialy valuable electromagnetic counterparts to
gravitational wave observations.  However, unless the black hole in these systems is of sufficiently low mass and sufficiently
high spin, the neutron star will simply be swallowed whole.  Nonetheless, even when
a neutron star survives a merger with a black hole intact,
there is still a promising source for a bright electromagnetic
event that has not previously been discussed.  In this work, we will show that when a black hole and neutron star are close enough
that the black hole orbits within the neutron-star magnetosphere, and the magnetic field threads the black-hole horizon, a
circuit is established with the neutron star as an external resistor,
the magnetic field lines as wires, the magnetospheric electron-positron plasma
providing current, and the black hole acting
as a battery.  The result, as we will show, is a $\gamma$-ray burst of very short duration which should nonetheless be detectable at
cosmological distances with sufficient timing resolution.  Since this process does not depend on the disruption of the neutron star
to operate, these events may prove more common than short GRBs generated by tidal disruption and accretion.

In the well-known Blandford-Znajek scenario \cite{Blandford:1977ds} for
driving electromagnetic jets in active galactic nuclei,
the spinning supermassive black hole at the center of a galaxy is embedded in a magnetic field.  The field is assumed to be
anchored in an accretion disk, which is further assumed to be an excellent conductor, so that the field lines are affixed to
the disk.  The spin of the black hole then twists the magnetic field lines, driving a Poynting flux at the expense of its own spin.
Recently, it has been shown \cite{Palenzuela:2010nf,Palenzuela:mag,merev} that this picture
could also be applied to a black hole-black hole binary, orbiting through a fixed background magnetic field, 
where the orbital energy drives the Poynting
flux.  However, it is unclear how magnetic fields as strong as is necessary ($B\approx 10^4$ G) can be expected to surround
an orbiting binary, since the binary will exert a
gravitational torque that counteracts the viscous inflow of the accretion disk
\cite{Artymowicz}, thereby inhibiting the deposition of a magnetic
field on the black hole.  It is conceivable
that plasma from the disk that does manage to accrete onto an orbiting binary
will tow its magnetic field along, but it seems less likely that this will cause the magnetic
field around the orbiting binary to be comparable in strength to the field found in the bulk of the accretion disk.

A different but related scenario, which will be the focus of this work, is the Poynting flux generated by the inspiral of a black hole-neutron star
binary.  Much attention has been given to the possibility that the neutron star will disrupt, and the remnant gas will be accreted
(see \cite{Lattimer} for the theoretical description, and \cite{Rosswog,Shibata,Foucart} 
for results from recent simulations).  However,
since neutron stars can possess surface magnetic fields of $\mathcal{O}(10^{12}$ G) (and $\sim 10^{15}$ G in the case of magnetars), it
is conceivable that they could provide a mechanism for driving a substantial Poynting flux.  Since the neutron star core that anchors
the magnetic field is superconducting,
the field lines will be affixed to the neutron star, and the relative motion and/or rotation of the neutron star
and the orbiting black hole will generate a Poynting flux.  Furthermore, since
we know that neutron stars can possess such large magnetic fields, the inspiral and eventual merger of black hole-neutron star binaries requires
no \emph{ad hoc} mechanism for transporting a sufficiently strong magnetic field to the vicinity of the black hole.  It is worth noting
that the magnetic fields of neutron stars may decay on shorter timescales ($\sim 10^7$ years \cite{Ostriker}) 
than the typical inspiral timescale ($10^9$ years \cite{Misner73}), so that the neutron star might have a smaller magnetic field
of $\mathcal{O}(10^{8}$ G) at merger, which would significantly mitigate the effect we describe.  
However, the decay timescale quoted above assumes
there is no replenishment of the magnetic field, for example through dynamo action, 
and that the field is entirely the result of amplification of the progenitor
star's field during the collapse to a neutron star.  Furthermore, it is possible that the tidal deformation of the neutron star 
during the late inspiral could result
in significant magnetic field amplification \cite{Rosswog}.
Even in the most pessimistic scenario, provided the neutron star magnetic field exceeds $10^8$ G at merger, 
the plasma will still be channeled to the magnetic poles
by the magnetic field \cite{Basko}, and the mechanism we describe will still operate, albeit at lower magnitudes.

To simplify the calculation, we employ the membrane paradigm
\cite{MembraneParadigm}, wherein the black hole event horizon is reinterpreted
as a two-dimensional conducting membrane, and the standard 
Maxwell equations in three dimensions govern the dynamics of electromagnetic
fields evolving along a sequence of hypersurfaces with minimal modification.  
Using this approach, the Blandford-Znajek effect for a single spinning
black hole immersed in plasma can be calculated
using simple circuit equations.  In that setting, the black hole spin drives a current along the magnetic field lines by Faraday induction.  
For the case we are addressing, Faraday induction results from a combination of the orbital
motion and spin of the black hole through the dipole magnetic field of the neutron star.  As the inspiral due to gravitational wave emission is
adiabatic, we can also treat the change in magnetic field strength threading the black hole adiabatically,
which makes the calculation surprisingly simple.  Both of these assumptions will break down at some point prior to merger, so our main
purpose will be to suggest this novel mechanism, present admittedly rough quantitative estimates, and motivate further investigation
via numerical simulations.  In particular, the dynamics of the magnetic field, which we are ignoring, is likely to become significant
and interesting at some point prior to merger, and could have a substantial quantitative impact on the observed luminosities.  

Throughout, we will use BH or NS to designate black hole or neutron star quantities, 
and $M$ and $r$ will refer to the mass and orbital separation of the system.
We will focus on the late inspiral regime, where
the gravitational waves are detectable and the magnetic field threading the black hole will be significant.
This regime will be well within the neutron-star light cylinder, 
defined by $r_c\leq c/\Omega_{\rm NS}$, where $\Omega_{\rm NS}$ is the spin frequency of
the neutron star, as well as the Alfv\'{e}n radius, where the magnetic field dominates the plasma dynamics.  
Therefore, the black hole orbits within the neutron star magnetosphere, where a tenuous plasma and the magnetic field
lines that thread it co-rotate with the neutron star \cite{pulsar}.
We focus on typical, slowly spinning neutron stars with spin periods of $\mathcal{O}$(1 sec),
so that the co-rotating magnetic field is well-approximated by a dipole rotating at the neutron-star spin frequency.
Assuming that the plasma in the neutron star magnetosphere is low-density and the force-free approximation applies \cite{Blandford:1977ds,pulsar},
and the magnetic field strength within the orbital plane is given by $|\vec{B}|=B_{\rm o}(r_{\rm NS}/r)^3$ corresponding to a typical dipole
(where $r_{\rm NS}$ and $B_{\rm o}$ are the radius and surface magnetic field strength of the neutron star), 
there will be a substantial electromotive force (emf) between the horizon of
the black hole and the poles of the neutron star.  This emf will
drive a current along magnetic field lines,
creating an astronomical circuit, where the black hole acts like a
battery that absorbs power due to its intrinsic resistivity and drives electromagnetic power toward the neutron star.

To estimate the power generated by the black hole-neutron star
circuit, we begin with Faraday's law around a black hole:
\beq
 \oint \alpha d\vec{l}\cdot\vec{E} = -\frac{d}{dt}\int \vec{B} \cdot \vec{dA} - \oint \alpha d\vec{l}\cdot(\vec{v}\times \vec{B}) 
\label{eq:faraday}
\eeq
where $\alpha$ accounts for the gravitational redshift of fiducial observers ($\alpha \equiv \sqrt{1-2M_{\rm BH}/r}$ for a nonspinning
black hole, where $M_{\rm BH}$ is the black-hole mass)
and $t$ is the time as measured by an observeer at rest at infinity. Of course, the
binary does not yield
a pure single black-hole spacetime but rather a spacetime better approximated by a
post-Newtonian expansion with two bodies as sources.
However, for our estimate we proceed with relativistic values
for a single black hole to gauge the magnitude of the effect.

\begin{figure}
\includegraphics[trim = 0mm 0mm 0mm 0mm, clip, scale=.30, angle=0]{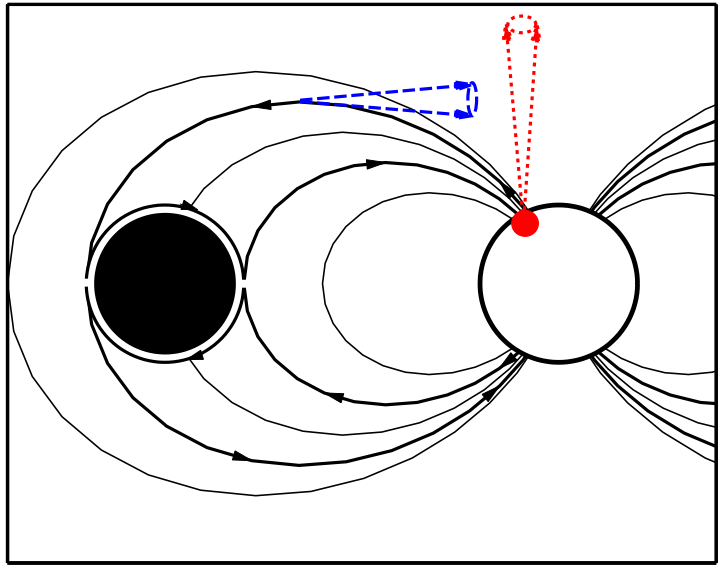}
\caption
{A black hole-neutron star binary, where the black hole (black) is threaded by the dipole magnetic field of the neutron star (white).  
The thick lines with arrows
show the integration contours used in Eq.~\eqref{eq:pot}, including the segment that runs over the horizon, just outside the black hole
\cite{MembraneParadigm}.  The Poynting flux from the black hole imparts kinetic energy into the plasma, which
in turn emits synchrotron radiation, and the magnetic field geometry accelerates the plasma and generates curvature radiation.
The synchrotron/curvature radiation will be relativistically beamed (blue dashed cone).  The plasma that reaches the neutron star surface
will induce radiation as its remaining kinetic energy dissipates by heating the surface.  This radiation will be nearly blackbody,
and will likely be relativistically beamed as well (red dotted cone).}
\label{fig:cartoon}
\end{figure}

Faraday's law has the usual interpretation. An emf along a closed loop
is related to the magnetic flux through an area enclosed by the loop
and to the motion of the circuit relative to
the magnetic field. While the law applies to any hypothetical loop we
might draw, we are interested in the emf along the
magnetic field lines, since they act as the wires
for conduction of the plasma.
The contour integration for employing Faraday's law in Eq.~\eqref{eq:faraday}
spans one hemisphere of the black-hole horizon, and connects the black hole's equator with one magnetic pole of the neutron star by running
parallel to the surrounding magnetic field lines (see
Fig.~\ref{fig:cartoon}). Assuming the rate of change of magnetic flux
is negligible for this contour on orbital timescales, the emf is due solely to the
relative motion of the contour through the
neutron-star magnetic field. 
This contour yields no contribution from the segments that run parallel to the magnetic field, so that only the contribution
from near the black-hole horizon remains.  The emf, or equivalently the 
potential difference across either black-hole hemisphere, 
is given by
\bea
V &=& \oint \alpha d\vec{l}\cdot\vec{E} = -\oint \alpha d\vec{l}\cdot(\vec{v}\times \vec{B}) \nonumber \\
&=& -2 r_{\rm H} B_{\rm o} \left(\frac{r_{\rm NS}}{r}\right)^3 \left[r\left(\Omega_{\rm orb}-\Omega_{\rm NS}\right)-\frac{a}{4\sqrt{2}}\right] \,
\label{eq:pot}
\eea
where $a\equiv S_{\rm BH}/M_{\rm BH}^2$ and $r_{\rm H}$ are the
dimensionless spin parameter and horizon radius of the black hole ($0\leq a\leq 1$, and $r_{\rm H}=2M$ for a nonspinning black hole)
$v=[r(\Omega_{\rm orb}-\Omega_{\rm NS})-a/(4\sqrt{2})]/\alpha$ is the azimuthal velocity of magnetic field lines as measured by non-inertial
observers at rest with respect to the black hole (\ie the fiducial
observers, or ``FIDOs'', of the membrane paradigm
\cite{MembraneParadigm}).  
$\Omega_{\rm orb}$ and $\Omega_{\rm NS}$ are the angular frequencies of the orbit and the neutron star spin, respectively,
with $\Omega_{\rm NS}$ defined here to be positive for spin aligned with the orbital angular momentum, and negative for anti-aligned spins.
Since the timescale for tidal locking greatly exceeds the inspiral timescale \cite{Bildsten:1992my},
realistic systems will not be tidally locked, and $v$ will be generically non-zero even for nonspinning black holes.
The term $a/(4\sqrt{2})$ is an approximation for the velocity of FIDOs as viewed from infinity due to the black-hole spin, and bears further
explanation.  For a spinning black hole, the angular velocity of FIDOs depends on polar angle because the horizon is no longer spherical.
However, for our purposes it is adequate to multiply the rms of the horizon radius,
$r_{\rm H}/\sqrt{2}$, by the angular frequency relative to FIDOs of field lines around a single Kerr black hole, $\Omega_{\rm BH}=a/(4r_{\rm H})$,
so that we will recover the standard Blandford-Znajek effect for $\Omega_{\rm orb}=\Omega_{\rm NS}=0$.

The emf drives a current provided by the plasma so that the circuit
obeys $V-IR_{\rm H}-IR_{\rm NS}=0$ giving $I=V/(R_{\rm H}+R_{\rm NS})$ where $R_{\rm H}$ is the
resistivity of the black hole membrane and $R_{\rm NS}$ is the resistivity
of the neutron star. 
We use the remarkable result that the horizon has an effective resistance of 377 $\Omega$ (or 4$\pi$ in geometrized units) 
\cite{Damour_1979}.
As $R_{\rm H}$ is simply the resistance of free space, we are assuming throughout that the resistance due to the magnetosphere
is negligible, and the resistance resulting from passing through the neutron star crust is
comparable to $R_{\rm H}$, so that equal amounts of power are absorbed by the black hole through Ohmic dissipation as is emitted
from near the horizon via a Poynting flux.  Clearly, this assumption depends on the composition of the crust and the dynamics within
the crust, which are very poorly constrained. 
We have also assumed that
the black hole horizon is sufficiently small and far away that the dipole field is approximately vertical as it threads the black hole.  This
approximation will break down for black holes much more massive than the neutron star, 
or at separations near merger, but we are not interested in systems
with very disparate masses ($M_{\rm BH} \gsim 500 \Msun$) since they will not be observable through gravitational waves with Advanced LIGO/Virgo
and their electromagnetic luminosity will be mitigated by having a large black-hole mass. 
Since most of our other approximations will also break down close to merger,
this should be a reasonable assumption.
 
We can now solve for the power generated by both hemispheres of the black-hole horizon:
\bea
L&=&2\frac{V^2}{R_{\rm H}}=\frac{8\epsilon}{\pi}(\alpha v)^2B^2M_{\rm BH}^2 \\
&=& 3\times 10^{46} \epsilon\left(\alpha \frac{v}{c}\right)^2\left(\frac{B_{\rm o}}{10^{12}\,{\rm G}}\right)^2\left(\frac{r_{\rm NS}}{r}\right)^6\left(\frac{M_{\rm BH}}{10\,\Msun}\right)^2\,{\rm erg/s} \,.\label{eq:powv}
\eea
We have assumed $\Omega_{\rm NS} \ll \Omega_{\rm orb}$, and we use 
$\epsilon \equiv (r_{\rm H}/2M)^2$ to account for the effect of black hole spin on the horizon size,
so that $\epsilon$ ranges from unity for a nonspinning black hole
to $1/4$ for maximal spin.
We choose 10 $\Msun$ as our fiducial black-hole mass
because, in the absence of any observational constraints for black hole-neutron star binaries,
we are left with results from population synthesis, which predict total mass distributions for these systems which peak at $\sim 8 \,\Msun$
\cite{Nelemans}, and measurements from low mass X-ray binaries, which show narrow distributions of $7.8\pm 1.2\,\Msun$ \cite{Ozel}.
The choice of a 10 $\Msun$ black hole paired with the standard choice of a 1.4 $\Msun$ neutron star
then has the virtue of being a conservative estimate, since $L\propto M_{\rm BH}^2r^{-6}\propsim M_{\rm BH}^{-4}$.

\begin{figure}
\includegraphics[trim = 0mm 0mm 0mm 0mm, clip, scale=.30, angle=0]{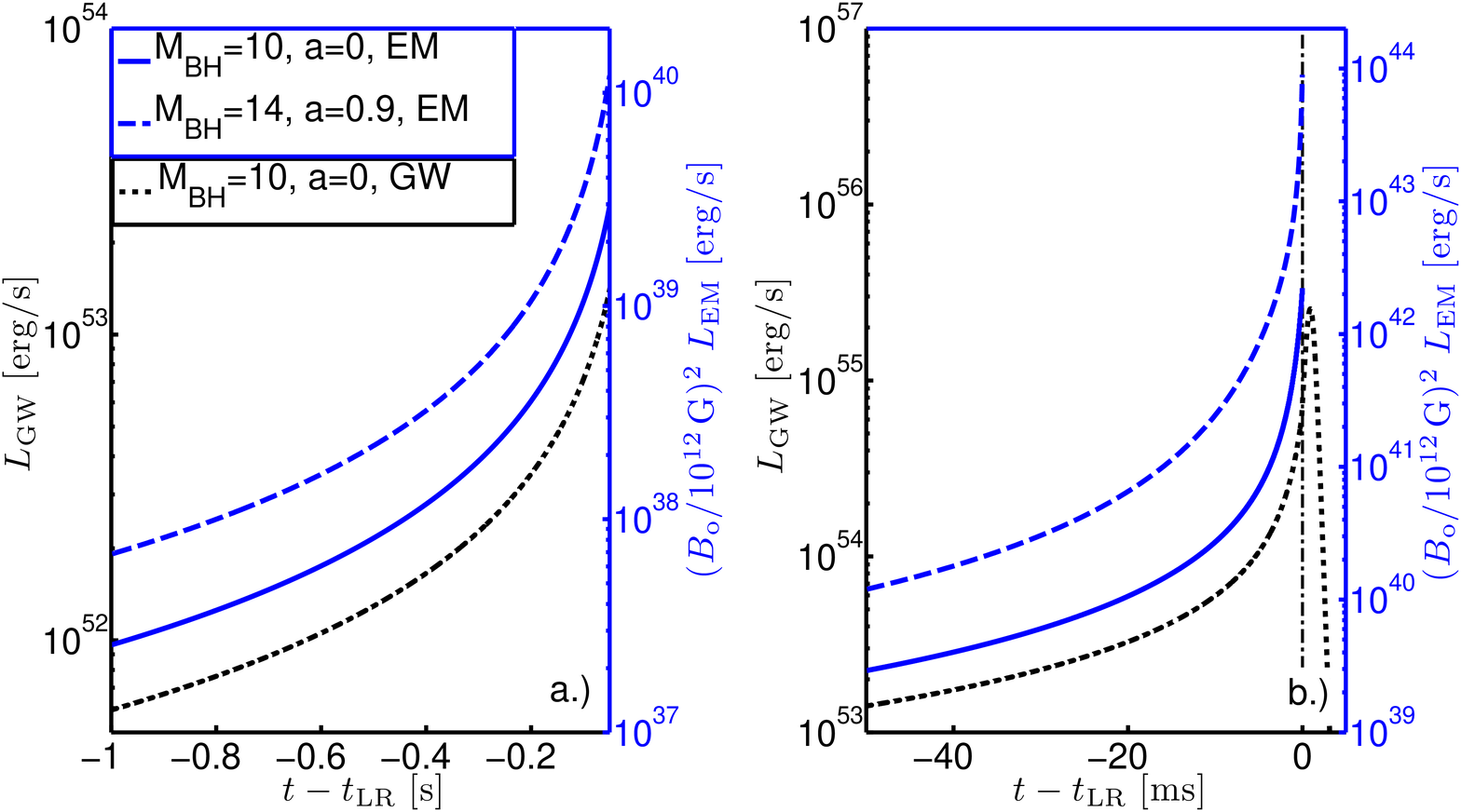}
\caption
{A comparison of the electromagnetic luminosity for our fiducial case (nonspinning black hole, $M_{\rm BH}=10\,\Msun$) and our most
luminous case ($a\equiv S/M_{\rm BH}^2=0.9$, $M_{\rm BH}=14\,\Msun$) with the gravitational wave luminosity for our fiducial case.
In (a.), we show the second preceding the final burst, which is the typical timescale for short GRBs.  In (b.), we show the final 50 ms of
the burst, which dominates the total energy output.  We note that the gravitational wave luminosity rises considerably after the light ring
(designated by a thin vertical dash-dotted line), so it seems likely that the peak electromagnetic emission from the process we describe may
substantially excede that predicted by Eqs.~\eqref{eq:powmax} and \eqref{eq:powmaxsp}, though numerical simulation will be required to
explore this regime.}
\label{fig:lum}
\end{figure}

Before proceeding, we must point out a few more caveats for applying Eq.~\eqref{eq:powv}.
We note that, if the neutron star spin period is $\mathcal{O}$(ms), rather than the far more common periods of $\mathcal{O}$(s),
or if the separation is large,
then the assumption that $\Omega_{\rm NS} \ll \Omega_{\rm orb}$ should not be employed, 
and the correct spin period should be used in Eq.~\eqref{eq:pot}. 
As millisecond pulsars generally have much weaker magnetic field strengths $B_{\rm o}\approx 10^{9}$ G, these systems are not only rarer, but also
far weaker than the systems of primary interest in this work.
In addition, for small stellar-mass black holes, the neutron-star radius or the tidal-disruption 
radius $r_{\rm tidal}\sim (M_{\rm BH}/M_{\rm NS})^{1/3}
r_{\rm NS}$, rather than the light-ring radius ($r=3M$ for a nonspinning black hole and $r=M$ for maximal spin),
may set the minimum distance for applicability of Eq.~\eqref{eq:powv}, as the calculation plainly will break down before the orbital separation
reaches these distances.  Finally, it must be emphasized that the energy source for the luminosity predicted by Eq.~\eqref{eq:powv} is a
combination of the black hole's spin energy and the gravitational
binding energy of the binary.  Therefore, we can expect no clear relationship between
Eq.~\eqref{eq:powv} and the spin-down rate of the black hole or neutron star in this scenario, as $L$ is non-vanishing 
(and, in fact, remains enormous) even in the limit that neither binary component is spinning.  
Furthermore, as the luminosity given by Eq.~\eqref{eq:powv} is
many orders of magnitude smaller than the gravitational wave luminosity for reasonable magnetic field strengths, 
the electromagnetic luminosity will not impact the dynamics.  In Fig.~\ref{fig:lum}, we show a comparison of the gravitational
wave luminosity with the electromagnetic luminosity for both nonspinning and highly spinning cases.

While Eq.~\eqref{eq:powv} will become increasingly inaccurate as the system approaches merger and magnetic field dynamics likely
become significant,
Eq.~\eqref{eq:powv} could yield a reasonable approximation at separations outside
the light ring if we
use an accurate trajectory informed by numerical relativity simulations.  For separations less than the radius of the light ring,
a significant fraction of the Poynting flux predicted by Eq.~\eqref{eq:powv} may be unable to escape from the black hole, so that the
peak luminosity is likely to occur at a separation near the light ring.  Therefore, in Fig.~\ref{fig:lum}, we calculate
the electromagnetic luminosity given by Eq.~\eqref{eq:powv},
with $v(t)$ calculated by evolving the pseudo-4PN Hamiltonian introduced in \cite{Buonanno:2007pf}
for the case of $a=0$, and by evolving
the Hamiltonian introduced in \cite{Barausse:2009xi} for the case of $a=0.9$.  The spinning black hole in our example has a mass
$M_{\rm BH}=14\,\Msun$, because this specific set of system parameters is not expected to result in the tidal disruption of the neutron
star \cite{Ferrari:2010}.  Since it is near the disruption threshold, it should be seen as a near-optimal case, although it may
nonetheless be more typical given the expected distribution of black hole spins.
We include the gravitational wave energy flux to third post-Newtonian order in the Taylor expansion
\cite{Blanchet:2004ek}.  We truncate the electromagnetic luminosity of both systems at their respective light rings, 
where we estimate the light ring location by finding the radius of a particle on a circular orbit with infinite momentum.
For comparison, we also calculate the gravitational wave luminosity for the nonspinning case
using the same inspiral model combined
with the implicit-rotating-source (IRS) merger model first presented in \cite{Baker:2008mj}.  As all of these models were tuned to
accurately match numerical relativity simulations of black hole-black hole binaries, they do not include any finite size effects from the neutron
star.  Although the models will therefore accumulate small differences from an actual simulation of the systems of interest, these errors
will be far smaller at late times than the other approximations we have described in deriving Eq.~\eqref{eq:powv}.  Naturally,
for sufficiently small black-hole masses and sufficiently large black-hole spins, the neutron star will be tidally disrupted prior to merger, 
so our approximations will fail sometime before the system reaches the disruption radius.  However, nonspinning black holes
with $M_{\rm BH}=3\,\Msun$ will not disrupt the neutron star 
prior to merger for realistic equations of state, and near-extremal spinning black holes will not
disrupt unless $M_{\rm BH} < 10\,M_{\rm NS}$ (Fig.~4 of \cite{Ferrari:2010}), so the process we describe may be much more common than
the signature from tidal disruption and subsequent accretion of the neutron star by the black hole.

We can use the results shown in Fig.~\ref{fig:lum} to estimate the peak luminosity for each system.  
Using the light-ring values for each case, we find 
\beq
L_{\rm peak}(a=0) \approx 2 \times 10^{42} \left(\frac{B_{\rm o}}{10^{12}\,{\rm G}}\right)^2\left(\frac{M_{\rm BH}}{10\,\Msun}\right)^{-4} \,{\rm erg/s}
\label{eq:powmax}
\eeq
for the nonspinning case, and
\beq
L_{\rm peak}(a=0.9) \approx 9 \times 10^{43} \left(\frac{B_{\rm o}}{10^{12}\,{\rm G}}\right)^2\left(\frac{M_{\rm BH}}{10\,\Msun}\right)^{-4} \,{\rm erg/s}  
\label{eq:powmaxsp}
\eeq
for the spinning case.
It is encouraging to note that Eq.~\eqref{eq:powmaxsp} is approaching 
the Blandford-Znajek luminosity of a single
maximally spinning black hole with the same total mass $M$, immersed in a magnetic field with the same strength $B_{\rm o}$ as our fiducial 
surface field.  This is
consistent with our expectation that the luminosity as $r\rightarrow M$ and $v_{\rm orb}\rightarrow c$ for the binary should equal the luminosity for a maximally-
spinning black hole, keeping all other variables equal, so that this observation is a useful sanity check.

While these estimates do not include any contribution from the final merger, it is still instructive to integrate Eq.~\eqref{eq:powv} over
the domain $r\geq r_{\rm LR}$, where $r_{\rm LR}$ is the light-ring radius, 
as a conservative estimate of the total energy emitted strictly by the process described herein, which yields
\beq
E_{\rm tot}(a=0) = \displaystyle\int_{r=r_{\rm LR}}^{\infty}\,dr\,\frac{L(r)}{dr/dt}\approx 3\times 10^{39}\left(\frac{B_{\rm o}}{10^{12}\,{\rm G}}\right)^2\left(\frac{M_{\rm BH}}{10\,\Msun}\right)^{-3} \,{\rm erg}
\label{eq:ener}
\eeq
for the nonspinning case, and
\beq
E_{\rm tot}(a=0.9) \approx 8\times 10^{41}\left(\frac{B_{\rm o}}{10^{12}\,{\rm G}}\right)^2\left(\frac{M_{\rm BH}}{10\,\Msun}\right)^{-3} \,{\rm erg}
\label{eq:enersp}
\eeq
for the spinning case,
where we again emphasize that this calculation is intended as an order-of-magnitude estimate, though it has the virtue of almost certainly being
an underestimate, as it neglects the merger. 
 
In the case of the Blandford-Znajek prototype, the system model would now be complete.  However, in the scenario presented here, the luminosity given
by Eq.~\eqref{eq:powv} is not simply emitted as a jet, as it would be for Blandford-Znajek or for 
orbiting black-hole binaries with a uniform magnetic field geometry.  Because the plasma in
the magnetosphere is constrained to follow magnetic field lines, it will be channeled from just outside the black-hole horizon to the neutron-star
magnetic poles.  As the plasma is accelerated forward by the Poynting flux and transversely by the curving field lines of the dipole geometry,
some fraction (which we will label $\eta$) of the power will be dissipated as synchrotron and 
curvature radiation that is strongly beamed by relativistic effects (see Fig.~\ref{fig:cartoon}).  Because the intensity
of the curvature radiation depends only
on the magnetic field geometry, and the intensity of the synchrotron radiation depends on the conserved magnetic flux, which determines the fraction
of electrons and positrons populating excited Landau levels, we expect the intensity of the emitted radiation 
as the plasma transits from the black hole to the
neutron star surface will be fairly uniform.  The synchrotron and curvature radiation will therefore uniformly sweep out the entire plane that contains
the black hole and neutron star and that is parallel to the magnetic field.  As this plane rotates with the orbiting binary, we have the interesting
phenomena of a beamed jet which covers the entire $4\pi$ sky each orbit.  Since the orbital timescale and the burst timescale (defined as the
time interval containing 90\% of the total energy) are both $\mathcal{O}$(ms), the entire sky should receive a brief blast of
luminosity at the magnitude of Eqs.~\eqref{eq:powmax} and \eqref{eq:powmaxsp}.
Judging from the behavior observed in numerical simulations of neutron star magnetospheres accelerating
a driven plasma wind \cite{cheng1},
the majority of the observed luminosity due to acceleration of the plasma within the magnetosphere will likely be due to curvature radiation,
which will dominate above 10 MeV, with synchrotron radiation accounting for the majority of observed radiation in hard X-rays and soft $\gamma$-rays,
with energies 10 keV $\lsim h \nu \lsim 10$ MeV.

As some unknown fraction of the emitted energy will still be in the form of plasma kinetic energy when the plasma reaches the neutron star,
the flux of plasma may induce a hot spot on the neutron-star surface (see Fig.~\ref{fig:cartoon}).  
Assuming even a small fraction of the luminosity predicted
by Eq.~\eqref{eq:powv} is so-deposited, the thermal blackbody emission re-radiated by the neutron star surface
will be substantially super-Eddington.  
Assuming a hot spot surface area $A\approx 1$ km$^2$, the temperature of the hot spot
will be given by
\beq
T = \left(\frac{L_{\rm BB}}{\sigma A}\right)^{1/4} 
= 1 \,{\rm MeV}\, [\epsilon (1-\eta)]^{1/4} \sqrt{\left(\alpha\frac{v}{c}\right) \left(\frac{B_{\rm o}}{10^{12}\,{\rm G}}\right) \left(\frac{r_{\rm NS}}{r}\right)^3 \left(\frac{M}{10\,\Msun}\right)}\,,
\label{eq:teff}
\eeq
where $L_{\rm BB}\equiv (1-\eta)L$ is the fraction of total luminosity (as given by Eq.~\eqref{eq:powv}) that is emitted from the neutron surface
and $\sigma$ is the Stefan-Boltzmann constant.
The maximum temperature of the hotspot, which results from inserting the maximum luminosities from 
Eqs.~\eqref{eq:powmax} and \eqref{eq:powmaxsp} into Eq.~\eqref{eq:teff},
is given by
\beq
T_{\rm max}(a=0)=100\,{\rm keV}\,(1-\eta)^{1/4}\sqrt{\frac{B_{\rm o}}{10^{12}\,{\rm G}}}\left(\frac{M}{10\,\Msun}\right)^{-1}\,,
\label{eq:tmax}
\eeq
for the nonspinning case, and
\beq
T_{\rm max}(a=0.9)=300\,{\rm keV}\,(1-\eta)^{1/4}\sqrt{\frac{B_{\rm o}}{10^{12}\,{\rm G}}}\left(\frac{M}{10\,\Msun}\right)^{-1}\,,
\label{eq:tmaxsp}
\eeq
for the spinning case.
Therefore, the total luminosity observed from this process will occur entirely within the hard X-ray--soft $\gamma$-ray bands.
This emission should therefore be observable by the Swift $\gamma$-ray burst and Fermi $\gamma$-ray space telescopes.
It is noteworthy that there is some evidence for a subclass of short
GRBs with durations less than 100 ms, photon energies near the upper bound of short GRBs, and no apparent afterglow \cite{vsgb}, 
which is quite consistent with
our expectations for the process we have described, and which is quite distinct from the expectations from neutron star-neutron star mergers
and the tidal disruption and subsequent accretion of neutron stars by low-mass black holes.
More speculatively, we note that these systems could bear some resemblance to the behavior observed in
the subset of soft $\gamma$-ray repeaters (SGRs) that display a regular sequence of pulses occurring
at $\mathcal{O}$(1 sec) intervals.
The leading candidate for driving SGRs is the onset of a star quake
on a magnetar \cite{Duncan}.  It is interesting to note, however, that if we take the observed SGR periodicity 
to be the orbital period, and assume $B_{\rm o} \approx 10^{15}$ G which is consistent with magnetars, then Eq.~\eqref{eq:powv}
predicts a luminosity of $\mathcal{O}(10^{35}$ erg/s, which
agrees very well with the observed luminosity from this subset of SGRs in quiescence \cite{mcgill}.  Clearly, this could simply
be an interesting coincidence.

Given the brevity of the signal, the Swift Burst Alert Telescope (BAT) \cite{swiftbat} 
or the Fermi GLAST Burst Monitor (GBM) \cite{fermigbm} would be the most
suitable instruments for observing these events.
The 100 $\mu$s and 2 $\mu$s timing resolutions of BAT and GBM, respectively, are more 
than adequate to resolve a signal with a duration of $\mathcal{O}$(ms), and their large fields-of-view are critical
for observing these transients.  We can use the flux sensitivities ($F_{\rm min}=0.07$ photons/cm$^2$/s for BAT and 0.7 photons/cm$^2$/s for GBM)
to estimate the detectable range for these events.  Using BAT and Eq.~\eqref{eq:powmax} and assuming the event
is beamed into a cone covering a solid angle $\Delta\Omega = 100\, {\rm deg}^2$, the maximum luminosity distance range
is given by 
\beq
D_L=\sqrt{\frac{L_{\rm peak}}{\Delta\Omega F_{\rm min}}}=60\sqrt{\frac{100\, {\rm deg}^2}{\Delta\Omega}}\, {\rm Mpc}\,.
\label{eq:dl}
\eeq
If we instead assume a narrower beaming $\Delta\Omega = 1\, {\rm deg}^2$, or assume a neutron-star surface magnetic field of $\mathcal{O}(10^{14}$ G),
the range would be 600 Mpc, or a redshift $z=0.1$ using WMAP seven-year
cosmological parameters \cite{wmap7}, which is comparable to the upper distance limit for 
gravitational wave detection of these systems with Advanced LIGO (see Fig.~4b of \cite{McWilliams:2010eq}).
As the estimates given by Eqs.~\eqref{eq:powmax} and \eqref{eq:powmaxsp} are likely to be exceedingly conservative estimates of the
true peak luminosity for this process (which will likely occur closer to merger), 
this justifies our claim that the process we describe could potentially drive an electromagnetic counterpart
to any gravitational wave signal from a black hole-neutron star binary that is observable by Advanced LIGO/Virgo.

Despite the uncertainties in the expected signature of these sources, we would expect that the electromagnetic emission would consist of a
synchrotron/curvature hard X-ray/soft $\gamma$-ray component
whose intensity varies over the orbital period, and a blackbody component with a peak energy corresponding to soft $\gamma$-rays,
whose intensity varies over an interval aprroximately given by the sum of 
the orbital period, neutron star spin period, and black hole spin period.
The total electromagnetic signature could therefore have two distinct observable components 
rising and falling with different periods if the source is close enough.  
Both the synchrotron/curvature and the blackbody emission will be relativistically beamed, which will increase the potential range for
observing these events, but may decrease the event rate for observations within a fixed volume.  If the synchrotron/curvature component
dominates, the beaming may not decrease the event rate, given that the unusual geometry of the system will blanket the entire
sky with radiation over intervals comparable to the burst duration. 
The numerical simulation of these systems could be done with existing
codes, and would be an invaluable confirmation (or refutation) of the mechanism which we have described.

\vspace{3mm}

\noindent {\bf Acknowledgements} 
We wish to thank Cole Miller, Frans Pretorius, and Anatoly Spitkovsky for helpful feedback on the manuscript.
This work was supported by NSF grant AST-0908365. JL gratefully acknowledges a KITP Scholarship.

\vspace{3mm}

\noindent {\bf Competing Interests}
The authors declare that they have no competing financial interests.

\vspace{3mm}

\noindent {\bf Author Information} 
Reprints and permissions information is available at www.nature.com/reprints. Correspondence
and requests for materials should be addressed to S.T.M.
(sean@astro.columbia.edu).

\end{document}